\documentclass[aps,prl,onecolumn,preprintnumbers,superscriptaddress,nofootinbib,floatfix,10pt]{revtex4}
\usepackage{amsfonts,amssymb,stmaryrd,latexsym,amsmath,braket}
\usepackage{subfigure,graphicx}

\begin{document} 

\title{Eigenvector continuation with subspace learning}

\author{Dillon~Frame}
\affiliation{Facility for Rare Isotope Beams and Department of Physics and Astronomy,
  Michigan State University, East Lansing, MI~48824, USA}
\affiliation{Department~of~Physics, North~Carolina~State~University, Raleigh,
NC~27695, USA}

\author{Rongzheng~He}
\affiliation{Facility for Rare Isotope Beams and Department of Physics and
Astronomy,
  Michigan State University, East Lansing, MI~48824, USA}
\affiliation{Department~of~Physics, North~Carolina~State~University, Raleigh,
NC~27695, USA}

\author{Ilse~Ipsen}
\affiliation{Department~of~Mathematics, North~Carolina~State~University, Raleigh, NC~27695, USA}

\author{Daniel~Lee}
\affiliation{Department of Electrical and Systems Engineering, Department of Computer and Information Science, Department of Bioengineering, University of Pennsylvania,
  Philadelphia,~PA~19104, USA}

\author{Dean~Lee}
\affiliation{Facility for Rare Isotope Beams and Department of Physics and
Astronomy,
  Michigan State University, East Lansing, MI~48824, USA}
\affiliation{Department~of~Physics, North~Carolina~State~University, Raleigh, NC~27695, USA}

\author{Ermal~Rrapaj}
\affiliation{Department~of~Physics, University~of~Guelph, Guelph, ON N1G
2W1, Canada}

\begin{abstract}
A common challenge faced in quantum physics is finding the extremal eigenvalues and eigenvectors of a Hamiltonian matrix in a vector space so large that
linear algebra operations on general vectors are not possible. There are numerous efficient methods developed for this task, but they generally fail when some control parameter in the Hamiltonian matrix exceeds some threshold value.    
In this work we present a new technique called eigenvector continuation that can extend the reach of these methods.  The key  insight is that while an eigenvector resides in a linear space with enormous dimensions, the eigenvector trajectory generated by smooth changes of the Hamiltonian matrix is well approximated by a very low-dimensional manifold.  We prove this statement using analytic
function theory and propose an algorithm to solve for the extremal eigenvectors.  We benchmark the method using several examples from quantum many-body theory.
\end{abstract}

\maketitle

We address the problem of finding the extremal eigenvalues and eigenvectors of a Hamiltonian matrix that is too large to store in computer memory.   This problem occurs regularly in quantum many-body theory and all existing methods either use Monte Carlo simulations, diagrammatic expansions, variational methods or some combination.  While these methods can be quite efficient, they can break down when one or more parameters in the Hamiltonian exceed
some tolerance threshold.  In Monte Carlo simulations the difficulty is caused by sign oscillations that cause positive and negative weights to cancel.  In diagrammatric expansions the problem is the divergence of the series expansion, and in variational methods the obstacle is capturing the details of the wave function using a variational ansatz or truncated basis expansion.  In this letter we introduce a new variational technique called eigenvector continuation that can be used to salvage the most difficult cases. 

In the mathematical literature, the terms eigenvector continuation
\cite{BDF08,BFGHK14,DWC14}, subspace tracking \cite{Saad14}, and successive constraint method for subspace
acceleration \cite{Sirkovic:2016} refer to the
computation of smoothly-varying bases for invariant subspaces of parameter-dependent
matrices. Although related, our approach is aimed
at determining eigenvalues and eigenvectors in a vector space so large that linear algebra operations on general vectors are not possible.  As a result, Krylov space methods as the Lanczos algorithm \cite{Lanczos:1950,Saadbook11} are not applicable in their usual formulation. Some examples of computational methods that  can tolerate
extremely
large-dimensional spaces are quantum Monte Carlo
simulations and many-body perturbation theory.
We assume that we have a computational method that can perform a limited set of operations such as inner products between eigenvectors of different Hamiltonian matrices and amplitudes of  eigenvectors sandwiching specific matrices such as a Hamiltonian matrix. In order to obtain results using only this limited information, we must be careful to maintain numerical accuracy and robustness in the presence of collinearities among the eigenvectors. 

In the following we demonstrate that when a control parameter in the Hamiltonian
matrix is varied smoothly, the extremal eigenvectors do not explore the large dimensionality of the linear space.  Instead they trace out trajectories with significant displacements in only a small number of linearly-independent directions.  We prove this statement using the principles of analytic continuation. Since the eigenvector trajectory is a low-dimensional manifold embedded in a very large space, we can ``learn" the eigenvector trajectory using data where the eigenvector is computable and apply eigenvector continuation to address problems where the computational method breaks down.    

Let us consider a finite-dimensional linear space and a  family of Hamiltonian matrices $H(c) = H_0 + cH_1$ where $H_0$
and $H_1$ are Hermitian. Let $\ket{\psi_j(c)}$ denote the eigenvectors of $H(c)$ with corresponding eigenvalues $E_j(c)$.   Since $H(c)$ is Hermitian for real $c$ and thus
diagonalizable,
 $E_j(c)$ has no singularities on the real axis, and we can define $\ket{\psi_j(c)}$
so that it also has no singularities on the real axis. We now  expand $\ket{\psi_j(c)}$ as a power series about the point $c=0$.  The series coefficients for $c^n$ are $\ket{\psi^{(n)}_j(0)}/n!$,
where the superscript $(n)$ denotes the $n^{\rm th}$ derivative.
An analogous series expansion can also be applied to the eigenvalue $E_j(c)$.  These
series converge for all
$|c|<|z|$, where $z$ and its complex conjugate $\bar{z}$ are the closest
singularities to $c=0$ in the complex plane.   In the following we discuss perturbation theory,
which can be regarded as the calculation of these series expansions in cases
where the eigenvalues and eigenvectors of $H_0$ are known or readily computable.

\begin{figure}[!ht]
\centering
\begin{subfigure}
\centering
\includegraphics[height=5cm]{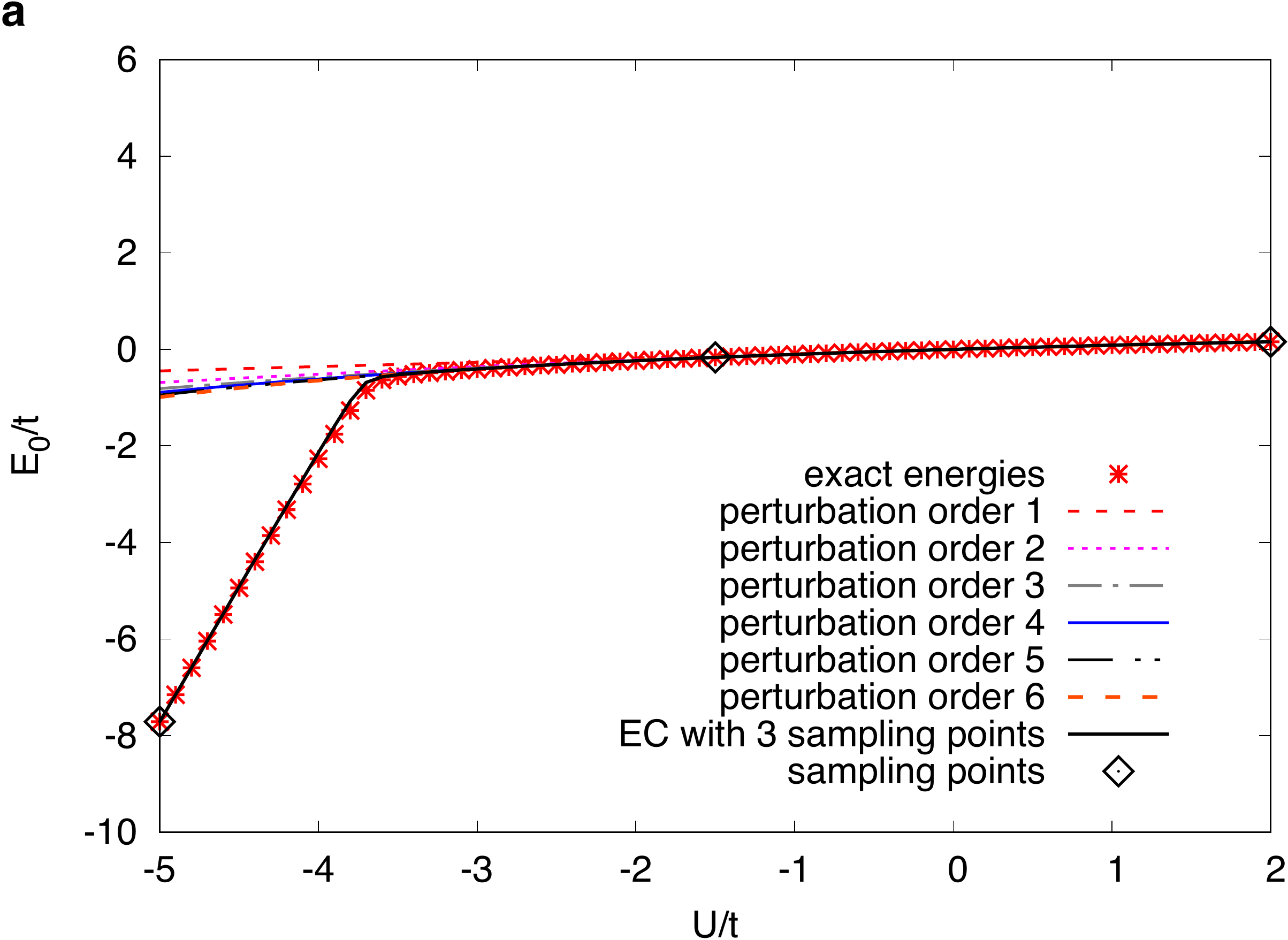}
\end{subfigure}
\begin{subfigure}
\centering
\includegraphics[height=5cm]{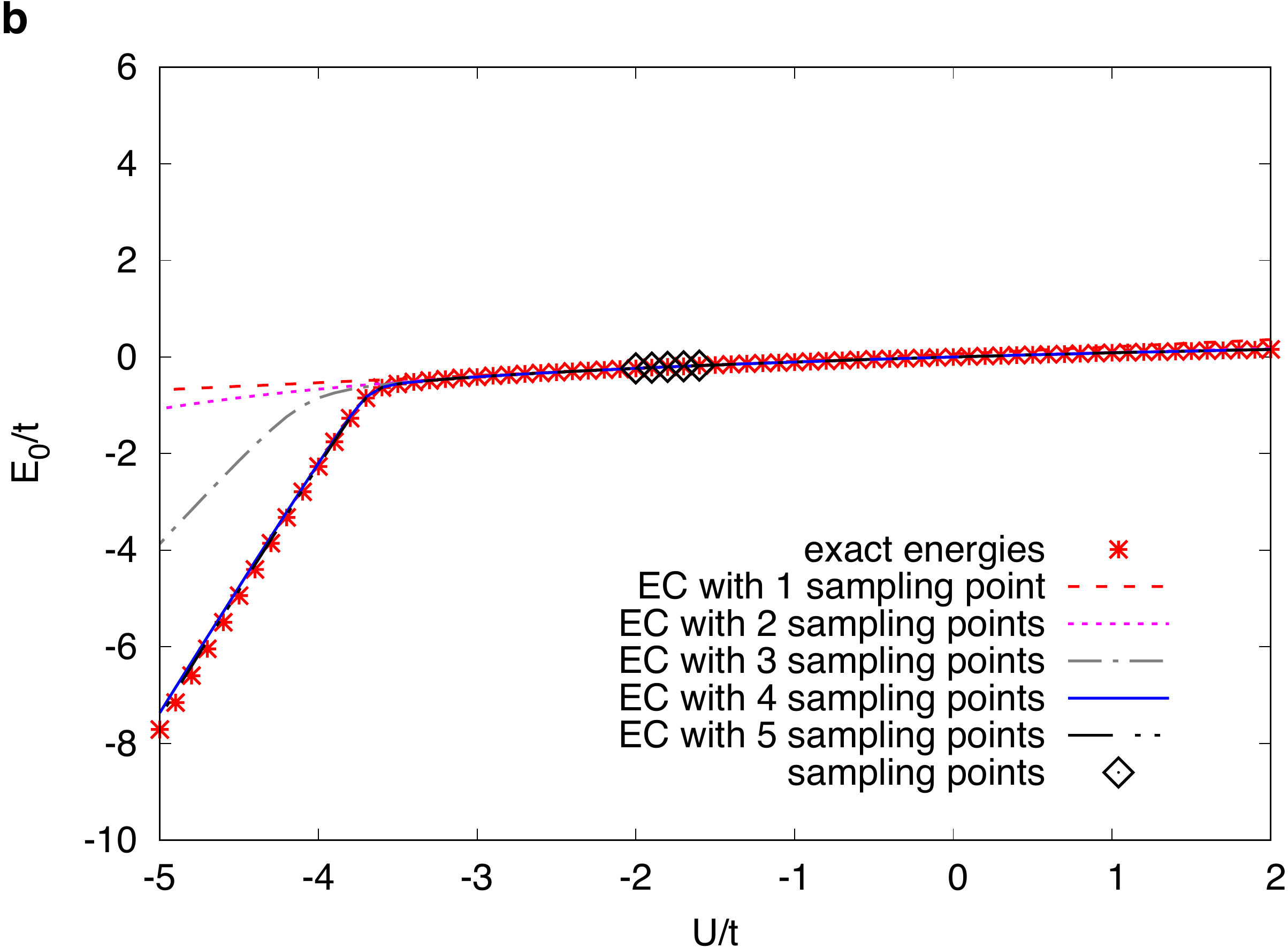}
\end{subfigure}
\caption{{\bf Ground state energy versus coupling.} In each panel {\bf a}
and {\bf b} we plot the ground state energy
$E_0/t$ versus coupling $U/t$ for four bosons
in the three-dimensional Bose-Hubbard model on a $4\times 4\times 4$ periodic
lattice. The red asterisks are the exact energies.
In panel {\bf a} the lines (1 = red dashed, 2 = magenta
dotted, 3 = grey dashed-dotted, 4
= blue solid, 5 = black
long-dashed-dotted, 6 = orange long-dashed) denote the first six orders of
the expansion of $E_0/t$
as a power series around the point $U/t = 0$. For comparison, the black solid
line shows $E_0/t$ computed using eigenvector continuation (EC) with the
three
sampling points shown as black diamonds. In panel {\bf b} the black diamonds
show the
five sampling points, and the lines (1 = red dashed, 2 = magenta
dotted, 3 = grey dashed-dotted, 4
= blue solid, 5 = black
long-dashed-dotted) denote the eigenvector continuation (EC) results for
$E_0/t$ when projecting onto 1, 2, 3, 4, or 5 vectors.}
\label{fourbody}
\end{figure}

In order to illuminate our discussion with a concrete example, we consider a quantum Hamiltonian known as the Bose-Hubbard model in three dimensions \cite{Gersch:1963a}.  It describes a system of identical bosons on a three-dimensional cubic lattice.  The Hamiltonian has a term proportional to $t$ that controls the lattice hopping of each boson, a term proportional
to $U$ that  controls the pairwise interactions between bosons on the same site, and a chemical potential $\mu$.  The full details of the model are given in Supplementary Materials.  We consider a system of four bosons with $\mu = -6t$ on a $4\times 4\times 4$ lattice.
We first try to use perturbation theory to compute the ground state energy eigenvalue $E_0$ in units of the hopping parameter $t$.  In panel {\bf a} of Fig.~\ref{fourbody} we show $E_0/t$ versus interaction strength $U/t$.  The red asterisks indicate the exact energies.
The lines (1 = red dashed, 2 = magenta
dotted, 3 = grey dashed-dotted, 4
= blue solid, 5 = black
long-dashed-dotted, 6 = orange long-dashed) denote the first six orders for the expansion of $E_0/t$
as a power series around $U/t = 0$.  We see that perturbation theory fails to converge when $U/t$ is less than about $-3.8$.\footnote{From the first six orders of the
expansion, the series might seem to converge to the wrong value for $U/t < -3.8$. However the series is divergent at higher orders.}  This is caused by branch point singularities at nearby points in the complex plane where the ground state eigenvalue is merging with another eigenvalue.

 The failure of perturbation theory is not surprising considering that the physical character of the ground state eigenvector changes significantly.  It is a Bose gas for $U/t > 0$, a weakly-bound state for $-3.8 < U/t < 0$, and then a tightly-bound cluster for $U/t < -3.8$. Although the eigenvector makes these changes in a linear space with hundreds of thousands of dimensions (before symmetrization), the eigenvector traces out a path with significant displacement in only a few independent directions.
To demonstrate this we compute the ground state eigenvectors at three sampling points, $U/t = -5.0, -1.5, 2.0$.  These three vectors span a three-dimensional subspace. We project the Hamiltonian for general $U/t$  onto this subspace and find the lowest eigenvalue and eigenvector. This technique is  an example of an approach we call eigenvector continuation.  In panel {\bf a} of Fig.~\ref{fourbody}
the black solid line shows $E_0/t$ computed using eigenvector continuation (EC) with the three sampling points shown as black diamonds. The agreement with the exact energies is quite good, and the same level of accuracy is found when comparing the eigenvector computed using EC to the exact eigenvector.

Eigenvector continuation can be used to ``learn'' sampling data from the region  $-3.8 < U/t
< 0$ and extrapolate to the regions $U/t < -3.8$ and $U/t
> 0$.
To demonstrate this we sample the ground state eigenvectors at five
points, $U/t = -2.0, -1.9, -1.8, -1.7, -1.6$. The results are shown in panel {\bf b} of Fig.~\ref{fourbody}. The red asterisks are the exact energies.
The black diamonds show the
five sampling points, and the lines (1 = red dashed, 2 = magenta
dotted, 3 = grey dashed-dotted, 4
= blue solid, 5 = black
long-dashed-dotted) denote the EC results for
$E_0/t$ when projecting onto 1, 2, 3, 4, or 5 vectors.  We see that the method converges rapidly and is able to capture the abrupt change in slope near $U/t = -3.8$.

  Why eigenvector continuation works and how fast it converges can be understood using analytic function theory. We return back to the series expansion for  $\ket{\psi_j(c)}$.  Although the series expansion about $c=0$ fails to converge for points $|c| > |z|$, we can define an analytic extension by constructing a new series about another point $c = w$, where $w$ is real and $|w| < |z|$. 
For this second series the coefficients of $(c-w)^n$ are $\ket{\psi^{(n)}_j(w)}/n!$.  We can  use the original series to express each $\ket{\psi^{(n)}_j(w)}$ in terms of $\ket{\psi^{(m)}_j(0)}$.  In this way we can approximate $\ket{\psi_j(c)}$ to arbitrary accuracy as a linear combination of the vectors $\ket{\psi^{(n)}_j(0)}$ in the region $|c-w| < |z-w|$ centered at $w$.
Using this process of analytic continuation repeatedly, we can reach any value of $c$ and express any $\ket{\psi_j(c)}$ to arbitrary accuracy as a linear combination of a finite number of vectors $\ket{\psi^{(n)}_j(0)}$.  The number of required vectors is determined by the number of different expansion centers needed in the analytic continuation and the rate of convergence
of each series expansion.  This explains why the trajectory traced out by $\ket{\psi_j(c)}$ moves in a small number of linearly-independent directions.

The basic strategy of eigenvector continuation is to ``learn'' the low-dimensional subspace that contains the eigenvector trajectory $\ket{\psi_j(c)}$.  We start with the lowest eigenvalue and
eigenvector in a given symmetry class. We then sample several values $c=c_i$ with $i=1,\cdots,K$ and compute the corresponding eigenvectors $\ket{\psi_j(c_i)}$.  The sampling values $c_i$ are chosen in the domain where
the computational method of choice is accurate.  The target
value $c = c_{\odot}$, where we want to determine $E_j(c_\odot)$ and $\ket{\psi_j(c_\odot)}$,
 will often lie in a region where direct calculation is no longer feasible. 
We then compute the inner products $N_{i',i}=\langle{\psi_j(c_{i'})} \vert \psi_j(c_i) \rangle$ and matrix elements $H_{i',i}=\langle{\psi_j(c_{i'})} \vert H(c_{\odot}) \vert \psi_j(c_i)
\rangle$ and solve
the generalized eigenvalue
problem.  This consists of finding the eigenvalues and eigenvectors of the $K$-dimensional matrix $N^{-1/2}HN^{-1/2}$, where $N^{-1/2}$ is the inverse square root of the positive matrix $N$.  For the lowest eigenvalue and eigenvector of each symmetry class, it suffices to compute the lowest eigenvalue and eigenvector of the $K$-dimensional matrix. We then proceed to the next-lowest eigenvalue and eigenvector in the symmetry class with the additional constraint that it is orthogonal to the lowest eigenvector.  Continuing on in this manner, any eigenvalue and eigenvector can in principle be calculated.
In cases where there are singularities near the real axis, the convergence of the method can be accelerated by including several eigenvectors $\ket{\psi_j(c_i)},\ket{\psi_
{j'}(c_i)},\cdots$ for each $c_i$. This procedure and the connection to Riemann sheets at branch point singularities is discussed in Supplementary Materials.

We now test the eigenvector continuation in a  many-body quantum Monte Carlo calculation.  We consider lattice simulations of pure neutron matter at leading order in chiral effective field theory.  Instead of using the lattice actions used in recent work \cite{Elhatisari:2016owd,Elhatisari:2017eno}, we purposely use the computationally difficult action described in Ref.~\cite{Lee:2016fhn}. Due to severe sign oscillations, it is not possible to do accurate simulations for more than four neutrons. Even extrapolation methods such as those discussed in Ref.~\cite{Alhassid:1993yd,Lahde:2015ona} provide no significant improvement due to the rapid onset of sign cancellations.
The leading-order action consists of the free neutron action, a single-site contact interaction between neutrons of opposite spins, and the two-body potential generated from the exchange of a pion.  This one-pion exchange potential is proportional to $g_A^2$, the square of the axial-vector coupling constant.  In contrast with the lattice actions used in 
 Ref.~\cite{Elhatisari:2016owd,Elhatisari:2017eno}, the short distance behavior of this one-pion exchange potential potential is not softened and, as a result, causes severe sign oscillations in the Monte Carlo simulations. We will consider the one-parameter family of lattice Hamiltonians $H(g_A^2)$ which results from varying $g_A^2$.  The desired target value of $g_A^2$ is the value $1.66$ used in Ref.~\cite{Lee:2016fhn}.  Details of the lattice action are presented in Supplementary Materials.

The
systems we calculate are the ground state energies of six and fourteen neutrons on a $4\times 4\times 4$ lattice with  spatial lattice spacing 1.97 fm and time lattice spacing 1.32 fm.  We are using natural units where $\hbar$ and the speed of light are set to $1$.  We use projection Monte Carlo with auxiliary fields to calculate the ground state energy.  Details of the simulation are presented in Supplementary Materials, and some reviews of the lattice methods can be found in Ref.~\cite{Lee:2008fa,Lee:2016fhn}.
 We first attempt to compute the ground state energies by direct calculation.  The errors are quite large due to sign oscillations.  For six neutrons the ground state energy is $E_0 = 12(^{+3}_{-4})$ MeV, and for fourteen neutrons  $E_0 = 42(^{+7}_{-15})$ MeV.
 
Next we use eigenvector continuation for the same systems with sampling data $g_A^2 = c_1,c_2,c_3$, where $c_1 = 0.25$, $c_2 = 0.60$, and $c_3 = 0.95$.  We use Monte Carlo simulations to calculate the ground state eigenvectors for $c_1,c_2,c_3$. In Table \ref{ec_table} we show the EC results using just one of the three vectors, two of the vectors, or all three vectors.
The error bars are\ estimates of the stochastic error and extrapolation error when taking the limit of infinite projection time.  For comparison we also show the direct calculation results.  We see that the EC results converge quite rapidly with the number of vectors included.  The results are also consistent with the direct calculation results, though with an error bar that is smaller by an order of magnitude.
\begin{table}[htbp]
\centering
\begin{tabular}{ccccc}
\hline\hline
$g_A^2$ values
& $E_0 (N=6)$ [MeV] 
& $E_0 (N=14)$ [MeV]
\\
\hline
\rule{0pt}{1.2em}%
$c_1$ & $14.0(4)$ & $48.8(6)$ \\
$c_2$ & $13.7(4)$ & $48.5(7)$ \\
$c_3$ & $13.8(6)$ & $48.8(8)$ \\
$c_2,c_3$ & $13.7(4)$ & $48.4(7)$ \\
$c_3,c_1$ & $13.8(4)$ & $48.8(6)$ \\
$c_1,c_2$ & $13.7(4)$ & $48.4(7)$ \\
$c_1,c_2,c_3$ & $13.7(4)$ & $48.4(7)$ \\
\hline
direct calculation & $12(^{+3}_{-4})$  & $42(^{+7}_{-15})$ \\
\hline\hline

\end{tabular}
\caption{Eigenvector continuation results for the ground state energy for six and fourteen neutrons using sampling data $g_A^2 = c_1,c_2,c_3$,
where $c_1 = 0.25$, $c_2 = 0.60$, and $c_3 = 0.95$. \ For comparison we also show the direct calculation results.}
\label{ec_table}
\end{table}  
 
Our calculations demonstrate the potential value of eigenvector continuation for quantum Monte Carlo simulations.  One can
use eigenvector continuation for interactions that  produce sign oscillations or noisy Monte Carlo simulations.  Eigenvector continuation can also be used to significantly extend the convergence of perturbation theory, and this will be demonstrated in a forthcoming publication.  While eigenvector continuation would not improve a Lanczos calculation using a truncated basis with fixed dimensions, eigenvector continuation  can be used to extend the reach of techniques that remove basis truncation errors\cite{Lee:2000xn}. The method is expected to be particularly useful for bound state calculations. For continuum states one should consider all low-lying continuum states 
in a finite volume together rather than picking out one continuum eigenvector at a time. This can be done using a framework such as the adiabatic projection method \cite{Elhatisari:2015iga,Elhatisari:2016hby}, which constructs continuum states for all possible relative displacements between clusters.

If the inner products $N_{i',i}$
and matrix elements $H_{i',i}$ can be computed with sufficient accuracy, then any eigenvector problem can be solved in this manner.  However, there are practical limits to the accuracy one can achieve for any computational method, and this sets limits on how far eigenvector  continuation can be pushed. In future work we will discuss machine learning techniques for optimizing the eigenvector continuation process \cite{jolliffe1986principal,seung2000manifold,saul2006spectral}.  While we have emphasized the use of eigenvector continuation to perform extrapolations in the control parameter $c$, there are also fascinating quantum systems where interpolation is the most interesting question.  One example is the phenomenon known as ``BCS-BEC crossover'' in degenerate fermionic systems at large scattering length \cite{Giorgini:2008zz}.  There are variational wave functions that work very well for the weak-coupling BCS side, and other variational wave functions that accurately describe the strong-coupling BEC side.  Our results here suggest that the crossover transition can be well represented using linear combinations of the different variational wave functions.  In the same manner, eigenvector continuation could also be used to study shape phase transitions in atomic nuclei \cite{Iachello:2001ph}.\footnote{We are grateful for discussions on this topic with Mark Caprio.}\  

We look forward to seeing future applications of eigenvector continuation when paired with computational
methods such as quantum Monte Carlo simulations, many-body perturbation theory, and variational methods.  We anticipate that eigenvector continuation can serve as a new theoretical tool to study quantum correlations, BEC-BCS\ crossover, shape transitions, entanglement, geometric phases, and quantum phase transitions at finite volume. 

We acknowledge partial financial support
from the U.S. Department of Energy (DE-FG02-03ER41260),
National Science Foundation, Army Research Laboratory, Office of Naval Research,
Air Force Office of Scientific Research, Department of Transportation, XDATA
Program of the Defense Advanced Research Projects Agency administered through
the Air Force Research Laboratory (FA8750-12-C-0323), Natural Sciences and
Engineering Research Council of Canada, Canada Foundation
for Innovation, Early Researcher Award program of the
Ontario Ministry of Research, Innovation and Science.   The computational
resources 
were provided by Michigan State University, North Carolina State University,
SHARCNET, NERSC, and the J\"{u}lich Supercomputing Centre at Forschungszentrum
J\"{u}lich.

\pagebreak

\section*{Supplemental Material}

\subsection*{Eigenvector continuation near branch point singularities}
\begin{figure}[!ht]
\centering
\includegraphics[height=5cm,angle=0]{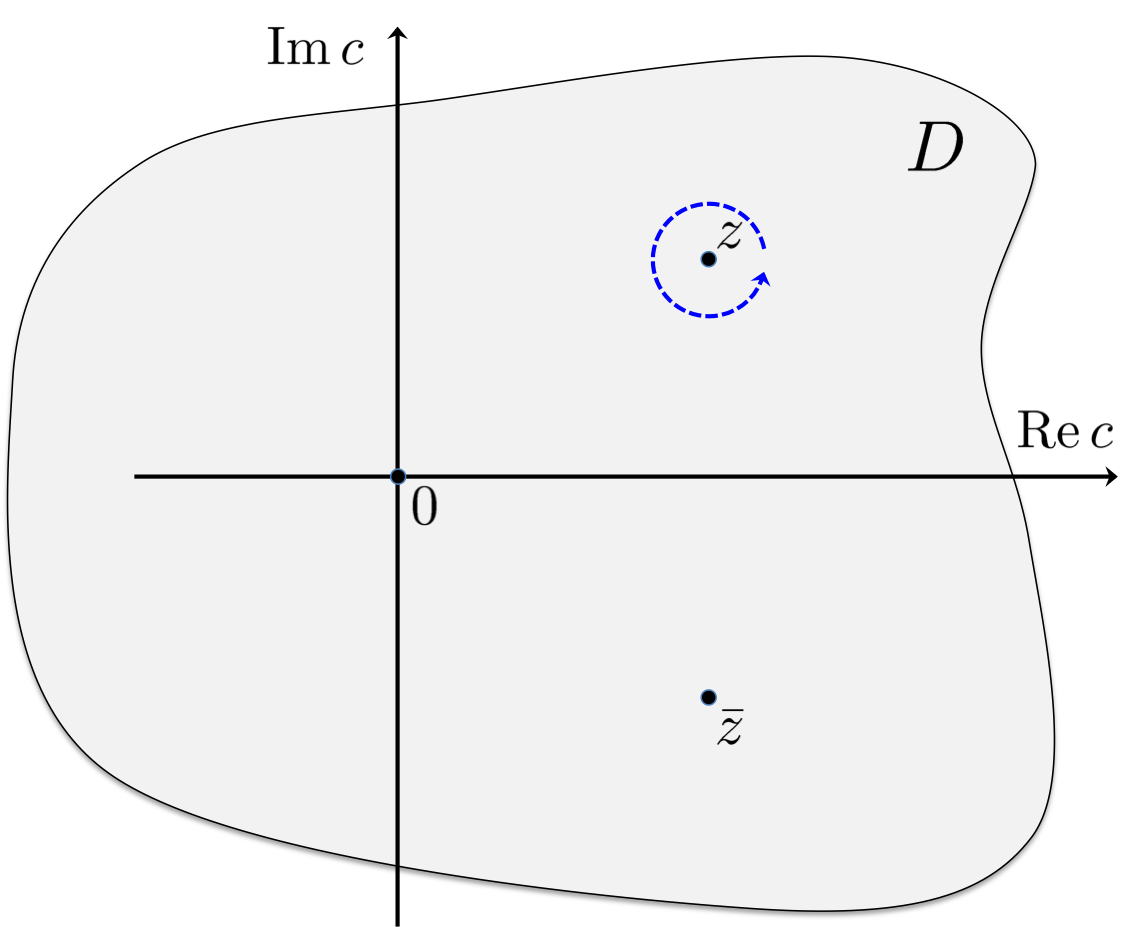}
\caption{{\bf Branch point singularities.} We show a finite region $D$ which
is portion of the Riemann surface
associated with $\ket{\psi_1(c)}$.  We assume that $\ket{\psi_1(c)}$ is analytic
everywhere in $D$ everywhere except possibly at the branch points $z$ and
$\bar{z}$.  The dashed blue arrow shows a small closed loop going counterclockwise
around the point $z$.}
\label{monodromy}
\end{figure}

As discussed in the main text, we consider a  family of Hamiltonian
matrices $H(c) = H_0 + cH_1$ where $H_0$
and $H_1$ are Hermitian. We write $\ket{\psi_j(c)}$ for the eigenvectors
of
$H(c)$ with corresponding eigenvalues $E_j(c)$.  We focus on one particular
eigenvector and eigenvalue which for simplicity we label as $\ket{\psi_1(c)}$
and $E_1(c)$ respectively.  Let $D$ be some finite region of the Riemann
surface associated with $\ket{\psi_1(c)}$.  We assume that $\ket{\psi_1(c)}$
is analytic everywhere in $D$ except possibly at branch point  singularities
located at $z$ and its complex conjugate $\bar{z}$. Since the elements of
the matrix $H(c)$
are analytic everywhere, the characteristic polynomial for $H(c)$ is also
analytic everywhere. We now consider what happens when traversing a closed
loop that starts and ends at the same
point
$c$ and which circles counterclockwise around the point $z$, as shown in
Fig.~\ref{monodromy}.  The result of this process on the eigenvalues and
eigenvectors of $H(c)$ will be called the monodromy transformation $T(z)$.

We first consider the case that $E_1(z)$ is a unique eigenvalue of $H(z)$.
 In this case $T(z)$ transforms the eigenvector $\ket{\psi_1(c)}$ to a vector
which is again proportional to $\ket{\psi_1(c)}$. In order to remove this
proportionality factor, we let $\ket{f}$ be some state in the linear space
such that $\langle
f \ket{\psi_1(c)} \ne 0$ for all points $c$ in $D$.  We  define the renormalized
eigenvector $\ket{\phi_1(c)}$ as
\begin{equation}
\ket{\phi_1(c)} = \ket{\psi_1(c)}/\langle f \ket{\psi_1(c)}.
\end{equation}Since $\langle f \ket{\phi_1(c)} = 1$ for all $c$ in $D$, $T(z)$
transforms $\langle f \ket{\phi_1(c)}$ into itself.  Therefore the vector
$\ket{\phi_1(c)}$ is also left invariant by $T(z)$.  While $\ket{\psi_1(c)}$
may have a branch point singularity at $z$, the renormalized eigenvector
$\ket{\phi_1(c)}$ is now analytic at $z$.   Since the subspace spanned by
the set of vectors $\ket{\phi_1(c_j)}$ is the same as the subspace spanned
by the set of vectors $\ket{\psi_1(c_j)}$, it is now clear that singularities
in the eigenvector normalization have no effect on the
convergence of the eigenvector continuation method. 

We now consider the case that $E_1(z)$ is not a unique eigenvalue for $H(z)$.
 In that case the monodromy transformation $T(z)$ can result in a permutation
of eigenvalues that are degenerate at $c=z$.  Without loss of generality
we consider the case that $T(z)$ induces a permutation cycle involving $k$
eigenvalues that we label as $E_1(c), \cdots, E_k(c)$.  With this notation
the monodromy transformation produces the cyclic permutation
\begin{equation}
T(z):E_1(c)\rightarrow E_2(c) \cdots \rightarrow E_k(c) \rightarrow E_1(c).
\end{equation}
The case $k=1$ will lead to the same conclusions as we found for the unique
eigenvalue $E_1(z)$, and so we focus on the case $k>1$.  Let us label the
image of $\ket{\phi_1(c)}$ under $T(z)$ as $\ket{\phi_2(c)}$, an eigenvector
with eigenvalue $E_2(c)$.  We continue labelling in this fashion so that
the monodromy transformation on the eigenvectors has the form
\begin{equation}
T(z):\ket{\phi_1(c)} \rightarrow \ket{\phi_2(c)} \cdots \rightarrow \ket{\phi_k(c)}
\rightarrow \ket{\phi_1(c)}.
\end{equation}
The statement that the last entry in this cycle is simply $\ket{\phi_1(c)}$
follows from the fact that $\langle f \ket{\phi_1(c)}$ is single-valued under
the   transformation $[T(z)]^k$.  

We note that the linear combinations
\begin{equation}
\ket{\gamma_n(c,z)} = \sum_{j=0}^{k-1}e^{i2\pi nj/k}\ket{\phi_j(c)} 
\end{equation}  
for $n = 0,\cdots k-1$ are eigenvectors of the monodromy transformation with
eigenvalues $e^{-i2\pi n/k}$.  We can now renormalize these vectors so that
the branch point singularity at $c=z$ is removed for each $n$,
\begin{equation}
\ket{\gamma'_n(c,z)} = (c-z)^{n/k}\ket{\gamma_n(c,z)}.
\end{equation}
These vectors $\ket{\gamma'_n(c,z)}$ are analytic at $c=z$. What we have
done is made linear combinations of the eigenvectors where all
non-integer fractional powers of $c-z$ have  been removed. In analogous fashion
we
can construct vectors $\ket{\gamma'_n(c,{\bar z})}$ that are analytic at
$c=\bar{z}$.  

Let us now consider performing eigenvector continuation with an EC subspace
consisting of  the $k$ eigenvectors $\ket{\psi_1(c_i)},$ $\cdots,$ $\ket{\psi_k(c_i)}$
for each sampling point $c_i$. We note that the singularities at $z$ and
$\bar{z}$ do not cause any convergence problems when determining $\ket{\psi_1(c)}$.
 This is because $\ket{\psi_1(c)}$ can be written as a linear combination
of vectors $\ket{\gamma'_n(c,z)}$ that are analytic at $z$, and, likewise,
 $\ket{\psi_1(c)}$ can be written as a linear
combination of vectors $\ket{\gamma'_n(c,\bar{z})}$ that are analytic at
$\bar{z}$. In order to accelerate the convergence near a branch point singularity,
the EC subspace should include all eigenvectors whose Riemann sheets become
entwined at the branch point.  

\subsection*{Bose-Hubbard model}
We consider an interacting system of identical bosons known as the three-dimensional
Bose-Hubbard model.
 The Hamiltonian has a term proportional to $t$ that controls the nearest-neighbor
hopping of each boson, a term proportional
to $U$ responsible for pairwise interactions between bosons on the
same site, and
a chemical potential $\mu$.  The Hamiltonian is defined on a three-dimensional
cubic lattice and has the form  
\begin{equation}
H = -t \sum_{\langle \bf{n'},\bf{n} \rangle} a^{\dagger}({\bf n'})a({\bf
n}) + 
\frac{U}{2}\sum_{\bf n}\rho(\bf{n})[\rho(\bf{n})-1]-\mu \sum_{\bf n}\rho(\bf{n}),
\end{equation}
where $a({\bf n})$ and $a^{\dagger}({\bf n})$ are the annihilation and creation
operators for bosons at lattice site ${\bf n}$, the first summation is over
nearest-neighbor pairs
$\langle {\bf n'},{\bf n}\rangle$, and $\rho({\bf n})$ is the density operator
$a^{\dagger}({\bf n})a(\bf{n})$.  In our calculations we consider a system
of four bosons with $\mu =
-6t$ on a $4\times 4\times 4$ lattice.  The choice $\mu =
-6t$ is convenient in that the ground state energy for the noninteracting
case $U=0$ has zero energy. 

\subsection*{Eigenvector continuation with two eigenvectors per sampling
point}
\begin{figure}[!ht]
\centering
\includegraphics[height=5cm]{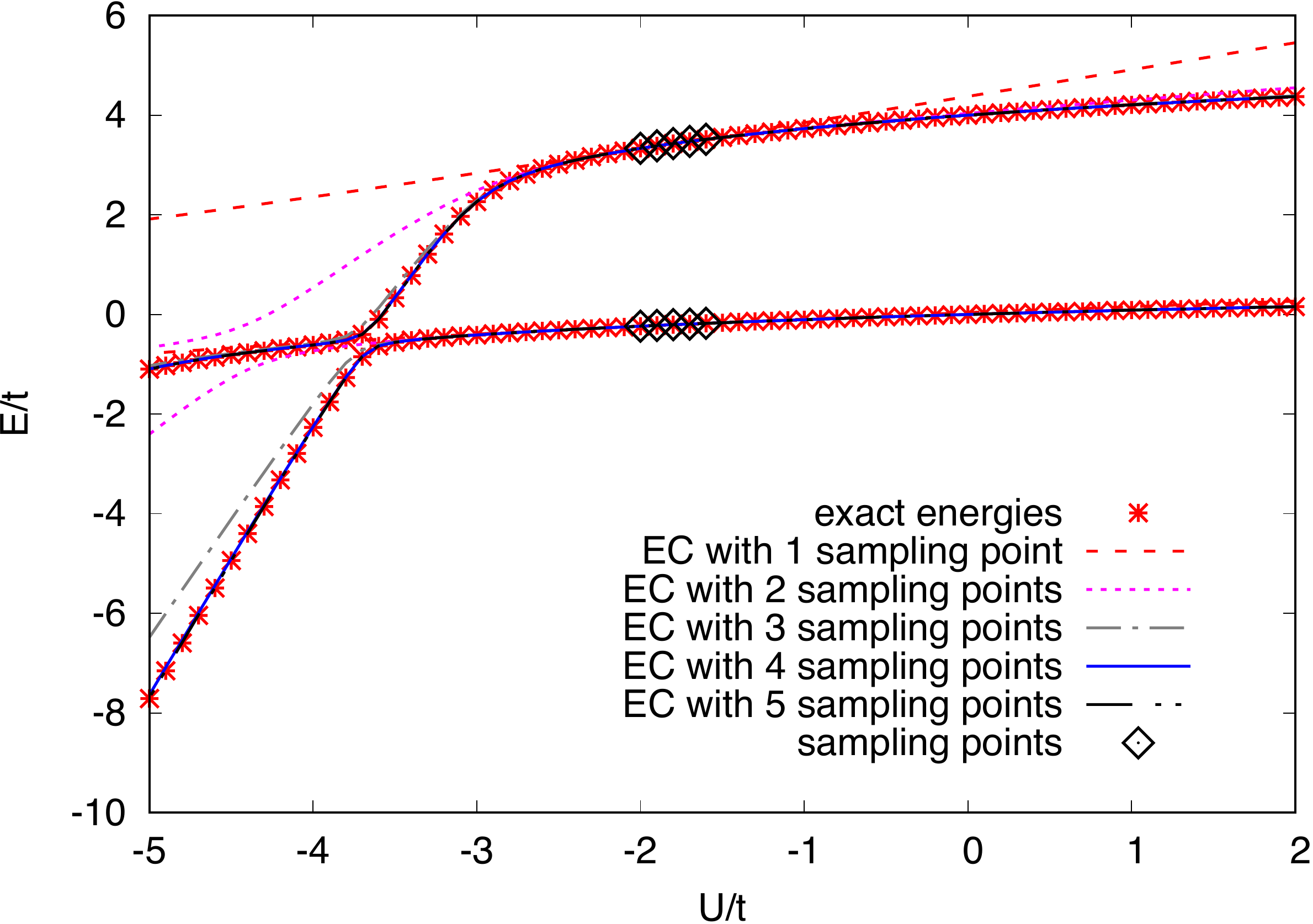}
\caption{{\bf Two lowest even-parity cubic-rotation-invariant energies versus
coupling.} We plot the two lowest even-parity cubic-rotation-invariant energies
$E/t$ versus coupling $U/t$ for four bosons
in the three-dimensional Bose-Hubbard model on a $4\times 4\times 4$ periodic
lattice. The red asterisks are the exact energies computed by sparse matrix
eigenvector iteration.  The black diamonds
show the
five sampling points, and the lines (1 = red dashed, 2 = magenta
dotted, 3 = grey dashed-dotted, 4
= blue solid, 5 = black
long-dashed-dotted) denote the eigenvector continuation (EC) results for
$E/t$ when projecting onto two eigenvectors per sampling point.}
\label{fourbody_2}
\end{figure}

In the main text we have demonstrated results for eigenvector continuation
while keeping one eigenvector of $H(c_i)$ per sampling point $c_i$.  But
we have also demonstrated that the convergence can be accelerated if the
EC subspace includes all eigenvectors whose Riemann sheets
become entwined at a nearby branch point singularity.  For real values of
the parameter $c$, the presence of nearby branch points can be seen as avoided
level crossings. In Fig.~\ref{fourbody_2} we plot the two lowest even-parity
cubic-rotation-invariant energies
$E/t$ versus coupling $U/t$ for four bosons
in the three-dimensional Bose-Hubbard model on a $4\times 4\times 4$ periodic
lattice. The red asterisks are the exact energies computed by sparse matrix
eigenvector iteration. We see clearly the avoided level crossing near $U/t
= -3.8$.

We now apply eigenvector continuation to this system once again, but this
time we keep the two lowest-energy even-parity
cubic-rotation-invariant eigenvectors for each of the five
sampling points, $U/t = -2.0, -1.9, -1.8, -1.7, -1.6$. The black diamonds
show the
five sampling points, and the lines (1 = red dashed, 2 = magenta
dotted, 3 = grey dashed-dotted, 4
= blue solid, 5 = black
long-dashed-dotted) denote the eigenvector continuation results for
$E/t$ when projected onto the EC subspace.
 We see that there is significant acceleration in the convergence to the
ground state energy when compared with keeping only one vector per sampling
point.  We note also that the excited state is accurately reproduced with
rapid convergence in the number of sampling points. 

\subsection*{Neutron matter lattice action}
We consider neutron matter as described by the leading-order lattice action
defined in Ref.~\cite{Lee:2016fhn}.  Since we are only considering neutrons,
we can simplify the lattice action by reducing the number of independent
contact interactions from two to one.
In the following we write $a_{i}({\bf n})$ and
$a_{i}^{\dagger}({\bf n})$ for the fermion annihilation and creation operators
with spin component $i$ at lattice site ${\bf n}$. The shorthand $a_{}({\bf
n})$ represents a column vector of nucleon components $a_{i}({\bf n})$, and
$a^{\dagger}({\bf n})$ represents a row vector of components $a^{\dagger}_{i}({\bf
n})$. We use
a spatial lattice spacing of
$1.97$ fm and time lattice spacing of 1.32 fm.  We express lattice quantities
in lattice units where physical quantities are multiplied by powers of the
spatial lattice spacing to produce dimensionless quantities.  We write $\alpha_t$
for the ratio $a_t/a$.
 
Our free non-relativistic lattice Hamiltonian
has the form\begin{equation}
H_{\rm free}(a^{\dagger},a) =\sum_{k=0,1}  \frac{(-1)^k}{2m}
\sum_{\bf n}\sum_{l=1,2,3} a^{\dagger}({\bf n}) \left[ a({\bf n}+k{\bf\hat{l}})
+ a({\bf n}-k{\bf \hat{l}})\right],
\end{equation} where ${\bf \hat{l}}$ denotes each of the three lattice unit
vectors, ${\bf \hat{1},{\bf \hat{2}},{\bf \hat{3}}}$. The nucleon mass is
taken to be $m=938.92$ MeV. We use the auxiliary-field formalism to produce
the interactions among neutrons. The Euclidean-time evolution operator over
$L_t$ time steps can be written as a product of transfer matrix operators
which depend on the
auxiliary field $s$ and pion field $\pi_0$,
\begin{equation}
U(L_t,g^2_A) = \int Ds D\pi_0 \; 
\exp{\left[-S_{ss}(s)
-S_{\pi_0\pi_0}(\pi_0)\right]}
 \left\{ M^{(L_t-1)}\cdots M^{(0)}\right\}.
\end{equation}  For later convenience we have explicitly listed the dependence
on the square of the axial-vector coupling $g_A$. The quadratic part of the
auxiliary-field action is 
\begin{equation}
S_{ss}(s) = \frac{1}{2} \sum_{{\bf n},n_t} s^{2}({\bf n},n_t).
\end{equation}
The quadratic part of the pion action is 
\begin{align}
S_{\pi_0\pi_0}(\pi_0)= & \frac{1}{2}\alpha_{t}m^2_{\pi} \sum_{{\bf n},n_t}\pi^2_0({\bf
n},n_t) \\
& +\frac{1}{2}\alpha_{t}%
\sum_{k=0,1}(-1)^k  \sum_{{\bf n},n_t} \sum_{l=1,2,3}\pi_0({\bf
n},n_t)
\left[ \pi_0({\bf n}+k{\bf \hat{l}},n_t) + \pi_0({\bf n}-k{\bf \hat{l}},n_t)
\right],
\end{align}
 where the pion mass is taken to be $m_\pi=134.98$
MeV. 

The auxiliary-field transfer matrix at time step $n_t$ is given by
\begin{equation}
M^{(n_t)}=\colon\exp\left[ -H^{(n_t)}(a^{\dagger},a,s,\pi_0)\alpha_{t}
\right]
\colon,
\end{equation}
where the $\colon \!\! \colon$ symbols indicate normal ordering, a rearrangment
of the operators with annihilation operators on the right and creation operators
on the left.  The Hamiltonian at time step $n_t$ has the form 
\begin{equation}
H^{(n_t)}(a^{\dagger},a,s,\pi_0)\alpha_{t}=H_{\text{free}}(a^{\dagger},a)\alpha_{t}+S^{(n_t)}_s(a^{\dagger},a,s)+S^{(n_t)}_{\pi_0}(a^{\dagger},a,\pi_0),
\end{equation}
where
\begin{align}
& S^{(n_t)}_{s}(a^{\dagger},a,s) =  \sqrt{-C\alpha_{t}} \sum_{\bf n} s({\bf
n},n_t)a^{\dagger}({\bf
n})a({\bf n}),  \\
& S^{(n_t)}_{\pi_0}(a^{\dagger},a,\pi_0) = \frac{g_A \alpha_t}{2 f_{\pi}}
\sum_{\bf n}\sum_{l=1,2,3}
\frac{1}{2}[\pi_0({\bf
n}+{\bf \hat{l}},n_t) - \pi_0({\bf
n}-{\bf \hat{l}},n_t)]a^{\dagger}({\bf
n})\sigma_l a({\bf n}).
\end{align}
Here $C$ is the coupling strength of the contact interaction, which we take
to be $-4.8\times 10^{-5} \, {\rm MeV}^{-2}$.  This is different from the
value used in Ref.~\cite{Lee:2016fhn} but is chosen to produce a more realistic
equation of state for neutron matter at the densities we probe.
$\sigma_l$ for $l=1,2,3$ denote Pauli matrices acting the neutron spins.
We
take the pion decay constant to be $f_\pi = 92.2\,{\rm MeV}$ and the axial-vector
coupling to be $g_A = 1.29$. Upon integrating over pion fields, one obtains
a one-pion exchange potential that is quadratic in $g_A$.

\subsection*{Neutron matter simulations}

We perform simulations of the ground state of six and fourteen neutrons in
a $4\times 4 \times 4$ box.  In physical units this is an $L = 7.9\,{\rm
fm}$ cubic box, and the corresponding number densities are $0.012$ fm$^{-3}$
for six neutrons and $0.028$ fm$^{-3}$
for fourteen neutrons.  We take our initial and final states to be a Fermi
gas of neutrons, which we write as $\ket{\Phi}$. We use Euclidean time projection
to produce the ground state for any chosen value of $g_A^2$.  We define $\ket{\Phi,n_t,g_A^2}$
as the projected state
\begin{equation}
\ket{\Phi,n_t,g_A^2} = U(n_t,g_A^2)\ket{\Phi}.
\end{equation} 
In the limit of large $n_t$, we obtain the ground state for the selected
value of $g_A^2$. 

We first present the results of the direct calculation without eigenvector
continuation.  In these calculations we  perform Monte Carlo simulations
to update the auxiliary and pion fields and  calculate the ratio
\begin{equation}
r(L_t) = \frac{\bra{\Phi}U(L_t,g_A^2)\ket{\Phi}}{\bra{\Phi}U(L_t-1,g_A^2)\ket{\Phi}}
\end{equation}
 in the limit of large $L_t$.
These ratios are converted into a measurement
of the energy using 
the relation 
\begin{equation}
E(L_t) = -\log[r(L_t)]/a_t.
\end{equation}
The ground state energy $E_0$ is determined using the asymptotic form\begin{equation}
E(L_t) \approx E_0 + c e^{-\delta E \cdot t}, \label{asymp}
\end{equation}  
where $t$ is the projection time, $t=L_t a_t$.
This asymptotic form takes into account the residual effects of excited state
contributions.

\begin{figure}[!ht]
\centering
\includegraphics[height=5cm]{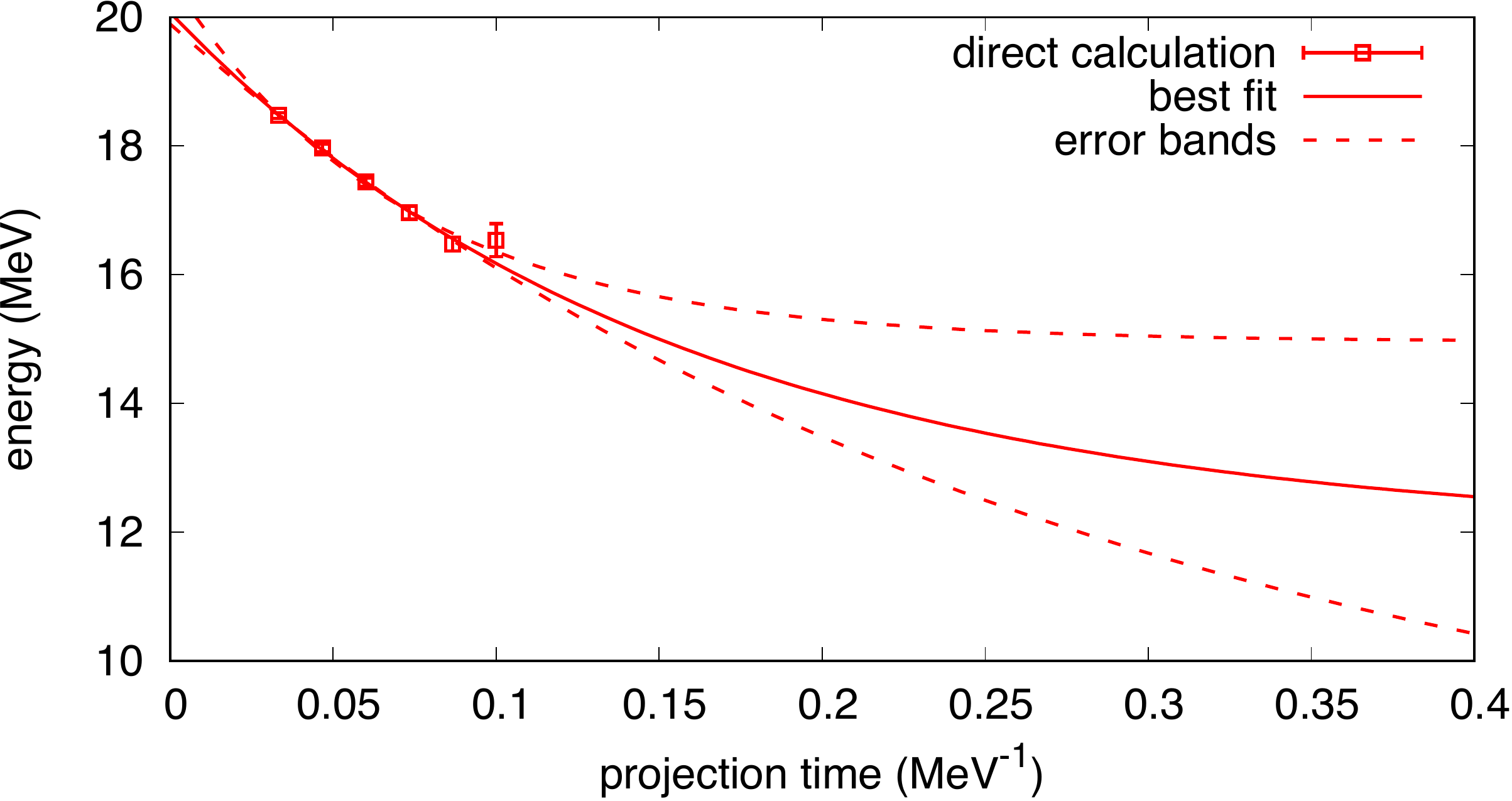}
\caption{{\bf Direct calculation for six neutrons.} Direct calculation of
the ground state energy versus projection
time for six neutrons.  The red open squares are the lattice results, including
one-standard-deviation error bars reflecting stochastic errors in the simulations.
 The best fit is shown as a red
solid line, and the 
limits of the one-standard-deviation error bands are shown as red dashed
lines.}
\label{neutrons6_direct}
\end{figure}

\begin{figure}[!ht]
\centering
\includegraphics[height=5cm]{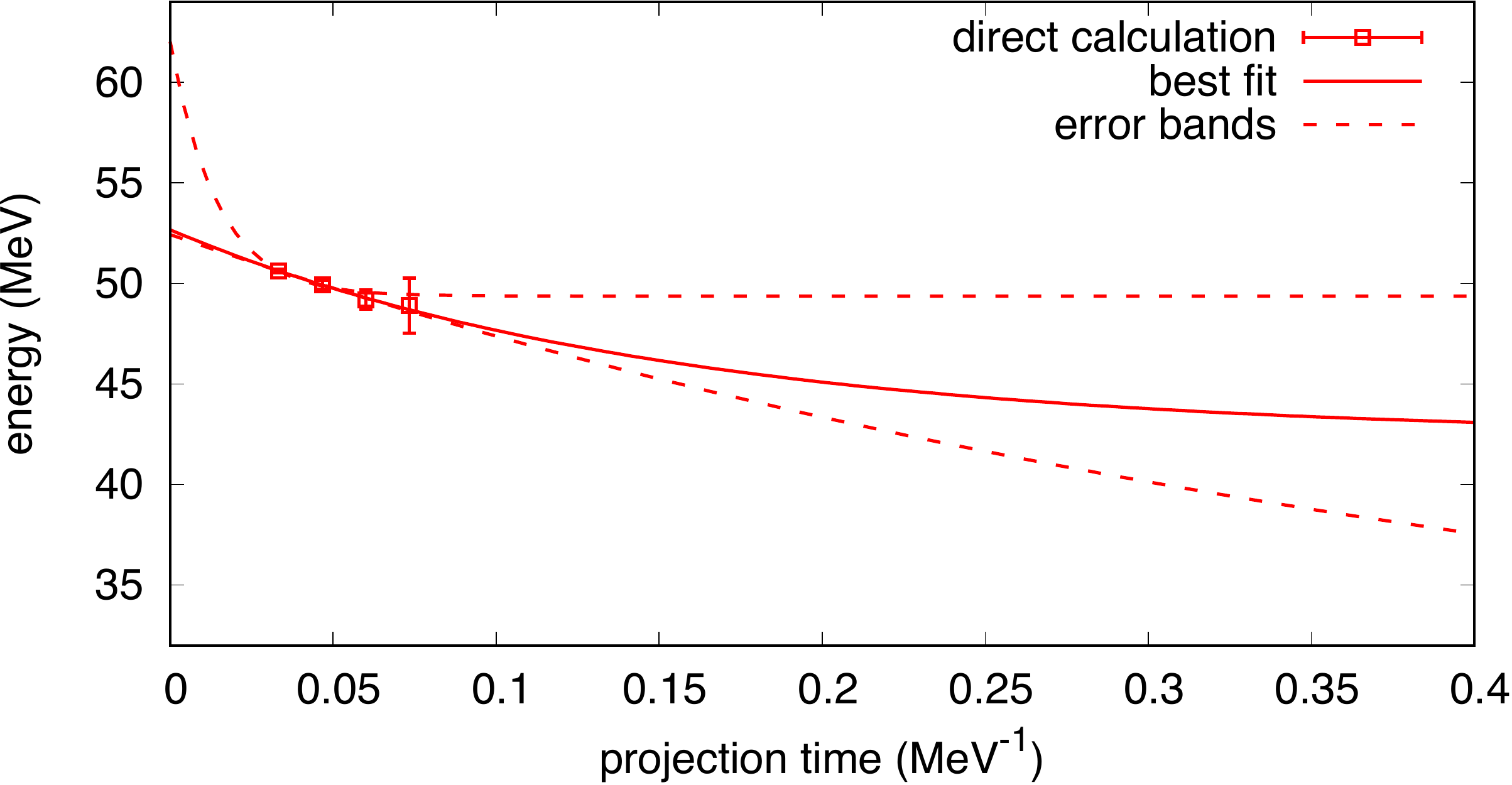}
\caption{{\bf Direct calculation for fourteen neutrons.} Direct calculation
of the ground state energy versus projection
time for fourteen neutrons.  The red open squares are the lattice results,
including
one-standard-deviation error bars reflecting stochastic errors in the simulations.
 The best fit is shown as a red
solid line, and the 
limits of the one-standard-deviation error bands are shown as red dashed
lines.}
\label{neutrons14_direct}
\end{figure}

In Fig.~\ref{neutrons6_direct} we show the results of the direct calculations
for six neutrons. The red open squares show the lattice results versus projection
time, including
one-standard-deviation error bars reflecting stochastic errors in the simulations.
 From the asymptotic form in Eq.~(\ref{asymp}), the best fit is shown as
a red
solid line, and the 
limits of the one-standard-deviation error bands are shown as red dashed
lines. 
In Fig.~\ref{neutrons14_direct} we show the results of the direct calculations
for fourteen neutrons. The red open squares show the lattice results versus
projection
time, including
one-standard-deviation error bars indicating stochastic errors.
 The best fit is shown as a red
solid line, and the 
limits of the one-standard-deviation error bands are shown as red dashed
lines. Due to large sign oscillations, it is not possible to do simulations
at large projection times.  This is reflected in the large uncertainties
of the ground state energy extrapolations.

For the eigenvector continuation calculations we compute the inner products
\begin{equation}
N_{i',i}(n_t) = \langle \Phi,n_t,c_{i'}  \vert \Phi,n_t,c_i \rangle
\end{equation}
for sampling values $g_A^2 = c_1,c_2,c_3$,
where $c_1 = 0.25$, $c_2 = 0.60$, and $c_3 = 0.95$. We also calculate the
matrix elements of the full transfer matrix (time evolution operator for
one time step) for the target value $g_A^2 = c_\odot=1.29^2 = 1.66$,
\begin{equation}
H_{i',i}(n_t) = \langle \Phi,n_t,c_{i'}  \vert U(1,c_\odot) \vert \Phi,n_t,c_i
\rangle.
\end{equation}
We solve the generalized eigenvalue 
value problem by diagonalizing the matrix $N^{-1/2}HN^{-1/2}$ and finding
the largest eigenvalue, $\lambda(n_t)$.  This is converted into a measurement
of the ground state energy using
the relation 
\begin{equation}
E(n_t) = -\log[\lambda(n_t)]/a_t.
\end{equation}  
We  then extrapolate to the limit $n_t \rightarrow \infty$ using the asymptotic
form \begin{equation}
E(n_t) \approx E_0 + c e^{-\delta E \cdot t}, 
\end{equation}  
where $t$ is the projection time $t=(2n_t+1) a_t$.

In these calculations we consider seven different choices for the EC subspace.
 For the first three we consider the one-dimensional subspaces spanned by
the single vector $\vert \Phi,n_t,c_i \rangle$ for all $i$. For the next
three we consider the two-dimensional subspaces spanned by two vectors $\vert
\Phi,n_t,c_i \rangle$ and $\vert \Phi,n_t,c_i' \rangle$ for all $i\ne i'$.
 Lastly we consider the full three-dimensional space spanned by $\vert \Phi,n_t,c_1
\rangle$, $\vert
\Phi,n_t,c_2 \rangle$, and $\vert \Phi,n_t,c_3 \rangle$.  

\begin{figure}[!ht]
\centering
\includegraphics[height=5cm]{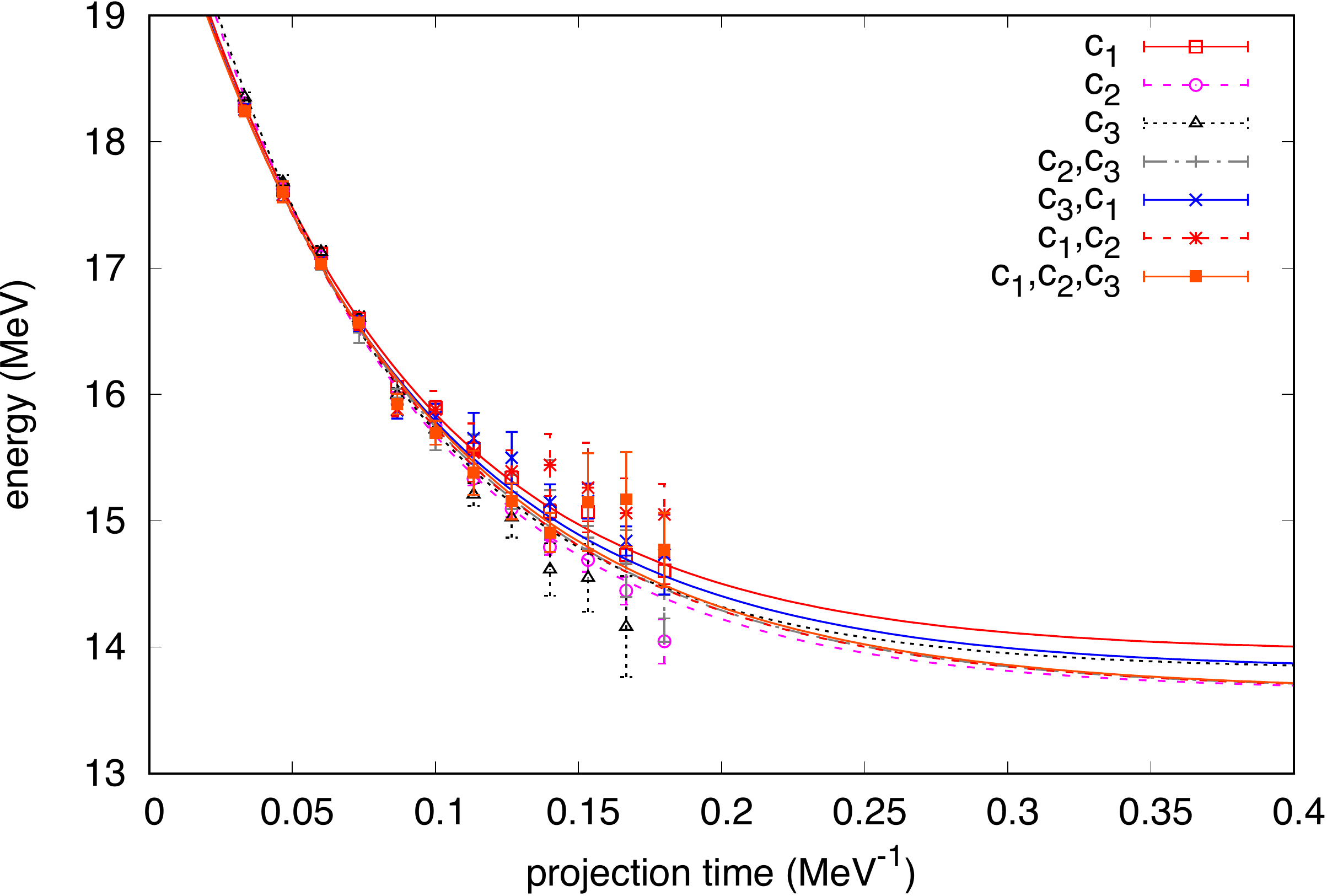}
\caption{{\bf Eigenvector continuation for six neutrons.} Eigenvector continuation
result for the ground state energy versus projection
time for six neutrons using sampling data $g_A^2 = c_1,c_2,c_3$, where $c_1
= 0.25$, $c_2 = 0.60$, and $c_3 = 0.95$.  We show results obtained using
one vector ($c_1$ = red open squares; $c_2$ = magenta open circles; $c_3$
= black open triangles), two vectors ($c_2,c_3$ = grey pluses; $c_3,c_1$
= blue crosses, $c_1,c_2$ = red asterisks), and all three vectors ($c_1,c_2,c_3$
= orange filled squares).
}
\label{neutrons6}
\end{figure}

The EC results for six neutrons are shown in Fig.~\ref{neutrons6}.  The
ground state energy
is shown versus projection
time using sampling data $g_A^2 = c_1,c_2,c_3$, where $c_1
= 0.25$, $c_2 = 0.60$, and $c_3 = 0.95$.  We show results obtained using
one vector, two vectors, and all three vectors.
The error bars are estimated using a jackknife analysis of the Monte Carlo
data.  We have imposed the variational constraint that the ground state energy
cannot increase as more vectors are included in the EC subspace. We find
that this technique makes a significant reduction in the extrapolation error.

\begin{figure}[!ht]
\centering
\includegraphics[height=5cm]{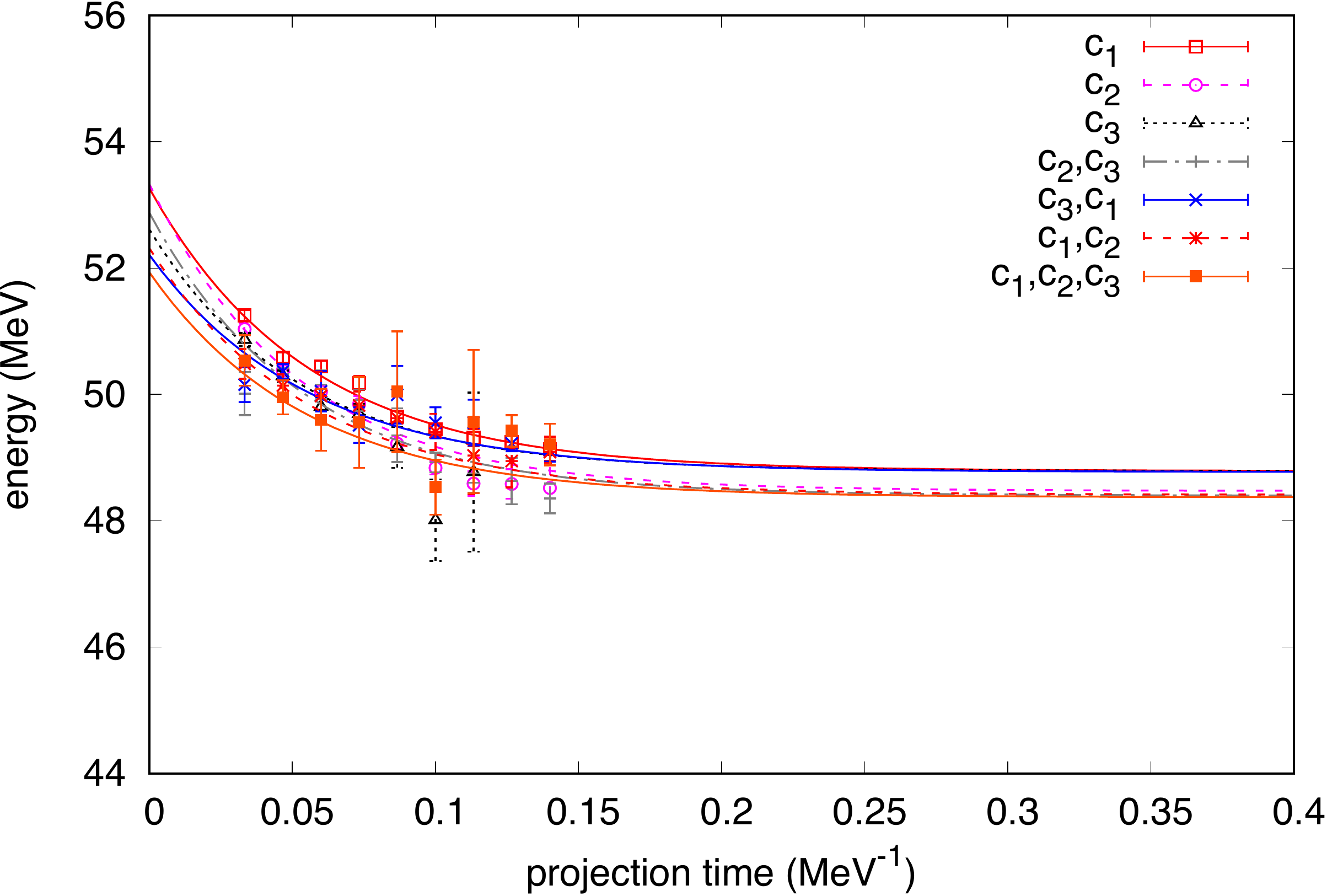}
\caption{{\bf Eigenvector continuation for fourteen neutrons.} Eigenvector
continuation results for the ground state energy
versus projection
time for fourteen neutrons using sampling data $g_A^2 = c_1,c_2,c_3$, where
$c_1
= 0.25$, $c_2 = 0.60$, and $c_3 = 0.95$.  We show results obtained using
one vector ($c_1$ = red open squares; $c_2$ = magenta open circles; $c_3$
= black open triangles), two vectors ($c_2,c_3$ = grey pluses; $c_3,c_1$
= blue crosses, $c_1,c_2$ = red asterisks), and all three vectors ($c_1,c_2,c_3$
= orange filled squares).
}
\label{neutrons14}
\end{figure}

The EC results for fourteen neutrons are shown in Fig.~\ref{neutrons14}.
 As before, the
ground state energy
is shown versus projection
time for sampling data $g_A^2 = c_1,c_2,c_3$, where $c_1
= 0.25$, $c_2 = 0.60$, and $c_3 = 0.95$.  Results are presented using
one vector, two vectors, and all three vector, and the error bars are estimated
using a jackknife analysis of the Monte Carlo
data.  We have
again imposed the variational
constraint that the ground state energy does not increase as more
vectors are included in the EC subspace.  

\bibliography{References}

\begin{thebibliography}{22}
\expandafter\ifx\csname natexlab\endcsname\relax\def\natexlab#1{#1}\fi
\expandafter\ifx\csname bibnamefont\endcsname\relax
  \def\bibnamefont#1{#1}\fi
\expandafter\ifx\csname bibfnamefont\endcsname\relax
  \def\bibfnamefont#1{#1}\fi
\expandafter\ifx\csname citenamefont\endcsname\relax
  \def\citenamefont#1{#1}\fi
\expandafter\ifx\csname url\endcsname\relax
  \def\url#1{\texttt{#1}}\fi
\expandafter\ifx\csname urlprefix\endcsname\relax\def\urlprefix{URL }\fi
\providecommand{\bibinfo}[2]{#2}
\providecommand{\eprint}[2][]{\url{#2}}

\bibitem[{\citenamefont{Bindel et~al.}(2008)\citenamefont{Bindel, Demmel, and
  Friedman}}]{BDF08}
\bibinfo{author}{\bibfnamefont{D.}~\bibnamefont{Bindel}},
  \bibinfo{author}{\bibfnamefont{J.}~\bibnamefont{Demmel}}, \bibnamefont{and}
  \bibinfo{author}{\bibfnamefont{M.}~\bibnamefont{Friedman}},
  \bibinfo{journal}{SIAM J. Sci. Comput.} \textbf{\bibinfo{volume}{30}},
  \bibinfo{pages}{637} (\bibinfo{year}{2008}).

\bibitem[{\citenamefont{Bindel et~al.}(2014)\citenamefont{Bindel, Friedman,
  Govaerts, Hughes, and Kuznetsov}}]{BFGHK14}
\bibinfo{author}{\bibfnamefont{D.}~\bibnamefont{Bindel}},
  \bibinfo{author}{\bibfnamefont{M.}~\bibnamefont{Friedman}},
  \bibinfo{author}{\bibfnamefont{W.}~\bibnamefont{Govaerts}},
  \bibinfo{author}{\bibfnamefont{J.}~\bibnamefont{Hughes}}, \bibnamefont{and}
  \bibinfo{author}{\bibfnamefont{Y.~A.} \bibnamefont{Kuznetsov}},
  \bibinfo{journal}{J. Comput. Appl. Math.} \textbf{\bibinfo{volume}{261}}
  (\bibinfo{year}{2014}).

\bibitem[{\citenamefont{Dijkstra et~al.}(2014)\citenamefont{Dijkstra, Wubs,
  Cliffe, and et~al.}}]{DWC14}
\bibinfo{author}{\bibfnamefont{H.~A.} \bibnamefont{Dijkstra}},
  \bibinfo{author}{\bibfnamefont{F.~W.} \bibnamefont{Wubs}},
  \bibinfo{author}{\bibfnamefont{A.~K.} \bibnamefont{Cliffe}},
  \bibnamefont{and} \bibinfo{author}{\bibnamefont{et~al.}},
  \bibinfo{journal}{Commun. Comput. Phys.} \textbf{\bibinfo{volume}{15}},
  \bibinfo{pages}{1} (\bibinfo{year}{2014}).

\bibitem[{\citenamefont{Saad}(2016)}]{Saad14}
\bibinfo{author}{\bibfnamefont{Y.}~\bibnamefont{Saad}}, \bibinfo{journal}{SIAM
  J. Matrix Anal. Appl.} \textbf{\bibinfo{volume}{37}}, \bibinfo{pages}{103}
  (\bibinfo{year}{2016}).

\bibitem[{\citenamefont{Sirkovic and Kressner}(2016)}]{Sirkovic:2016}
\bibinfo{author}{\bibfnamefont{P.}~\bibnamefont{Sirkovic}} \bibnamefont{and}
  \bibinfo{author}{\bibfnamefont{D.}~\bibnamefont{Kressner}},
  \bibinfo{journal}{SIAM J. Matrix Anal. Appl.} \textbf{\bibinfo{volume}{37}},
  \bibinfo{pages}{695} (\bibinfo{year}{2016}).

\bibitem[{\citenamefont{Lanczos}(1950)}]{Lanczos:1950}
\bibinfo{author}{\bibfnamefont{C.}~\bibnamefont{Lanczos}}, \bibinfo{journal}{J.
  Res. Nat. Bur. Stand.} \textbf{\bibinfo{volume}{45}}, \bibinfo{pages}{255}
  (\bibinfo{year}{1950}).

\bibitem[{\citenamefont{Saad}(2011)}]{Saadbook11}
\bibinfo{author}{\bibfnamefont{Y.}~\bibnamefont{Saad}},
  \emph{\bibinfo{title}{Numerical methods for large eigenvalue problems}},
  vol.~\bibinfo{volume}{66} of \emph{\bibinfo{series}{Classics in Applied
  Mathematics}} (\bibinfo{publisher}{Society for Industrial and Applied
  Mathematics (SIAM), Philadelphia, PA}, \bibinfo{year}{2011}),
  \bibinfo{note}{revised edition of the 1992 original}.

\bibitem[{\citenamefont{Gersch and Knollman}(1963)}]{Gersch:1963a}
\bibinfo{author}{\bibfnamefont{H.~A.} \bibnamefont{Gersch}} \bibnamefont{and}
  \bibinfo{author}{\bibfnamefont{G.~C.} \bibnamefont{Knollman}},
  \bibinfo{journal}{Phys. Rev.} \textbf{\bibinfo{volume}{129}},
  \bibinfo{pages}{959} (\bibinfo{year}{1963}).

\bibitem[{\citenamefont{Elhatisari
  et~al.}(2016{\natexlab{a}})}]{Elhatisari:2016owd}
\bibinfo{author}{\bibfnamefont{S.}~\bibnamefont{Elhatisari}}
  \bibnamefont{et~al.}, \bibinfo{journal}{Phys. Rev. Lett.}
  \textbf{\bibinfo{volume}{117}}, \bibinfo{pages}{132501}
  (\bibinfo{year}{2016}{\natexlab{a}}), \eprint{1602.04539}.

\bibitem[{\citenamefont{Elhatisari et~al.}(2017)\citenamefont{Elhatisari,
  Epelbaum, Krebs, L{\"a}hde, Lee, Li, Lu, Mei{\ss}ner, and
  Rupak}}]{Elhatisari:2017eno}
\bibinfo{author}{\bibfnamefont{S.}~\bibnamefont{Elhatisari}},
  \bibinfo{author}{\bibfnamefont{E.}~\bibnamefont{Epelbaum}},
  \bibinfo{author}{\bibfnamefont{H.}~\bibnamefont{Krebs}},
  \bibinfo{author}{\bibfnamefont{T.~A.} \bibnamefont{L{\"a}hde}},
  \bibinfo{author}{\bibfnamefont{D.}~\bibnamefont{Lee}},
  \bibinfo{author}{\bibfnamefont{N.}~\bibnamefont{Li}},
  \bibinfo{author}{\bibfnamefont{B.-N.} \bibnamefont{Lu}},
  \bibinfo{author}{\bibfnamefont{U.-G.} \bibnamefont{Mei{\ss}ner}},
  \bibnamefont{and} \bibinfo{author}{\bibfnamefont{G.}~\bibnamefont{Rupak}}
  (\bibinfo{year}{2017}), \eprint{1702.05177}.

\bibitem[{\citenamefont{Lee}(2017)}]{Lee:2016fhn}
\bibinfo{author}{\bibfnamefont{D.}~\bibnamefont{Lee}}, \bibinfo{journal}{Lect.
  Notes Phys.} \textbf{\bibinfo{volume}{936}}, \bibinfo{pages}{237}
  (\bibinfo{year}{2017}), \eprint{1609.00421}.

\bibitem[{\citenamefont{Alhassid et~al.}(1994)\citenamefont{Alhassid, Dean,
  Koonin, Lang, and Ormand}}]{Alhassid:1993yd}
\bibinfo{author}{\bibfnamefont{Y.}~\bibnamefont{Alhassid}},
  \bibinfo{author}{\bibfnamefont{D.}~\bibnamefont{Dean}},
  \bibinfo{author}{\bibfnamefont{S.}~\bibnamefont{Koonin}},
  \bibinfo{author}{\bibfnamefont{G.}~\bibnamefont{Lang}}, \bibnamefont{and}
  \bibinfo{author}{\bibfnamefont{W.}~\bibnamefont{Ormand}},
  \bibinfo{journal}{Phys. Rev. Lett.} \textbf{\bibinfo{volume}{72}},
  \bibinfo{pages}{613} (\bibinfo{year}{1994}), \eprint{nucl-th/9310026}.

\bibitem[{\citenamefont{L{\"a}hde et~al.}(2015)\citenamefont{L{\"a}hde, Luu,
  Lee, Mei{\ss}ner, Epelbaum, Krebs, and Rupak}}]{Lahde:2015ona}
\bibinfo{author}{\bibfnamefont{T.~A.} \bibnamefont{L{\"a}hde}},
  \bibinfo{author}{\bibfnamefont{T.}~\bibnamefont{Luu}},
  \bibinfo{author}{\bibfnamefont{D.}~\bibnamefont{Lee}},
  \bibinfo{author}{\bibfnamefont{U.-G.} \bibnamefont{Mei{\ss}ner}},
  \bibinfo{author}{\bibfnamefont{E.}~\bibnamefont{Epelbaum}},
  \bibinfo{author}{\bibfnamefont{H.}~\bibnamefont{Krebs}}, \bibnamefont{and}
  \bibinfo{author}{\bibfnamefont{G.}~\bibnamefont{Rupak}},
  \bibinfo{journal}{Eur. Phys. J.} \textbf{\bibinfo{volume}{A51}},
  \bibinfo{pages}{92} (\bibinfo{year}{2015}), \eprint{1502.06787}.

\bibitem[{\citenamefont{Lee}(2009)}]{Lee:2008fa}
\bibinfo{author}{\bibfnamefont{D.}~\bibnamefont{Lee}}, \bibinfo{journal}{Prog.
  Part. Nucl. Phys.} \textbf{\bibinfo{volume}{63}}, \bibinfo{pages}{117}
  (\bibinfo{year}{2009}), \eprint{0804.3501}.

\bibitem[{\citenamefont{Lee et~al.}(2001)\citenamefont{Lee, Salwen, and
  Windoloski}}]{Lee:2000xn}
\bibinfo{author}{\bibfnamefont{D.}~\bibnamefont{Lee}},
  \bibinfo{author}{\bibfnamefont{N.}~\bibnamefont{Salwen}}, \bibnamefont{and}
  \bibinfo{author}{\bibfnamefont{M.}~\bibnamefont{Windoloski}},
  \bibinfo{journal}{Phys. Lett.} \textbf{\bibinfo{volume}{B502}},
  \bibinfo{pages}{329} (\bibinfo{year}{2001}), \eprint{hep-lat/0010039}.

\bibitem[{\citenamefont{Elhatisari et~al.}(2015)\citenamefont{Elhatisari, Lee,
  Rupak, Epelbaum, Krebs, L{\"a}hde, Luu, and
  Mei{\ss}ner}}]{Elhatisari:2015iga}
\bibinfo{author}{\bibfnamefont{S.}~\bibnamefont{Elhatisari}},
  \bibinfo{author}{\bibfnamefont{D.}~\bibnamefont{Lee}},
  \bibinfo{author}{\bibfnamefont{G.}~\bibnamefont{Rupak}},
  \bibinfo{author}{\bibfnamefont{E.}~\bibnamefont{Epelbaum}},
  \bibinfo{author}{\bibfnamefont{H.}~\bibnamefont{Krebs}},
  \bibinfo{author}{\bibfnamefont{T.~A.} \bibnamefont{L{\"a}hde}},
  \bibinfo{author}{\bibfnamefont{T.}~\bibnamefont{Luu}}, \bibnamefont{and}
  \bibinfo{author}{\bibfnamefont{U.-G.} \bibnamefont{Mei{\ss}ner}},
  \bibinfo{journal}{Nature} \textbf{\bibinfo{volume}{528}},
  \bibinfo{pages}{111} (\bibinfo{year}{2015}), \eprint{1506.03513}.

\bibitem[{\citenamefont{Elhatisari
  et~al.}(2016{\natexlab{b}})\citenamefont{Elhatisari, Lee, Mei{\ss}ner, and
  Rupak}}]{Elhatisari:2016hby}
\bibinfo{author}{\bibfnamefont{S.}~\bibnamefont{Elhatisari}},
  \bibinfo{author}{\bibfnamefont{D.}~\bibnamefont{Lee}},
  \bibinfo{author}{\bibfnamefont{U.-G.} \bibnamefont{Mei{\ss}ner}},
  \bibnamefont{and} \bibinfo{author}{\bibfnamefont{G.}~\bibnamefont{Rupak}},
  \bibinfo{journal}{Eur. Phys. J.} \textbf{\bibinfo{volume}{A52}},
  \bibinfo{pages}{174} (\bibinfo{year}{2016}{\natexlab{b}}),
  \eprint{1603.02333}.

\bibitem[{\citenamefont{Jolliffe}(1986)}]{jolliffe1986principal}
\bibinfo{author}{\bibfnamefont{I.~T.} \bibnamefont{Jolliffe}}, in
  \emph{\bibinfo{booktitle}{Principal component analysis}}
  (\bibinfo{publisher}{Springer}, \bibinfo{year}{1986}), pp.
  \bibinfo{pages}{115--128}.

\bibitem[{\citenamefont{Seung and Lee}(2000)}]{seung2000manifold}
\bibinfo{author}{\bibfnamefont{H.~S.} \bibnamefont{Seung}} \bibnamefont{and}
  \bibinfo{author}{\bibfnamefont{D.~D.} \bibnamefont{Lee}},
  \bibinfo{journal}{Science} \textbf{\bibinfo{volume}{290}},
  \bibinfo{pages}{2268} (\bibinfo{year}{2000}).

\bibitem[{\citenamefont{Saul et~al.}(2006)\citenamefont{Saul, Weinberger, Ham,
  Sha, and Lee}}]{saul2006spectral}
\bibinfo{author}{\bibfnamefont{L.~K.} \bibnamefont{Saul}},
  \bibinfo{author}{\bibfnamefont{K.~Q.} \bibnamefont{Weinberger}},
  \bibinfo{author}{\bibfnamefont{J.~H.} \bibnamefont{Ham}},
  \bibinfo{author}{\bibfnamefont{F.}~\bibnamefont{Sha}}, \bibnamefont{and}
  \bibinfo{author}{\bibfnamefont{D.~D.} \bibnamefont{Lee}},
  \bibinfo{journal}{Semisupervised learning} pp. \bibinfo{pages}{293--308}
  (\bibinfo{year}{2006}).

\bibitem[{\citenamefont{Giorgini et~al.}(2008)\citenamefont{Giorgini,
  Pitaevskii, and Stringari}}]{Giorgini:2008zz}
\bibinfo{author}{\bibfnamefont{S.}~\bibnamefont{Giorgini}},
  \bibinfo{author}{\bibfnamefont{L.~P.} \bibnamefont{Pitaevskii}},
  \bibnamefont{and}
  \bibinfo{author}{\bibfnamefont{S.}~\bibnamefont{Stringari}},
  \bibinfo{journal}{Rev. Mod. Phys.} \textbf{\bibinfo{volume}{80}},
  \bibinfo{pages}{1215} (\bibinfo{year}{2008}), \eprint{0706.3360}.

\bibitem[{\citenamefont{Iachello}(2001)}]{Iachello:2001ph}
\bibinfo{author}{\bibfnamefont{F.}~\bibnamefont{Iachello}},
  \bibinfo{journal}{Phys. Rev. Lett.} \textbf{\bibinfo{volume}{87}},
  \bibinfo{pages}{052502} (\bibinfo{year}{2001}).

\end{thebibliography}

\bibliographystyle{apsrev}

\end{document}